# Ultrahigh stiffness and anisotropic Dirac cones in BeN$_4$ and MgN$_4$ monolayers: A first-principles study

Bohayra Mortazavi [a,*,#], Fazel Shojaei[b,**,#], Xiaoying Zhuang[a,c]

[a]Chair of Computational Science and Simulation Technology, Institute of Photonics, Department of Mathematics and Physics, Leibniz Universität Hannover, Appelstraße 11,30167 Hannover, Germany.
[b]Department of Chemistry, Faculty of Nano and Bioscience and Technology, Persian Gulf University, Bushehr 75169, Iran.
[c]College of Civil Engineering, Department of Geotechnical Engineering, Tongji University, 1239 Siping Road Shanghai, China.

## Abstract

Beryllium polynitrides, BeN$_4$ is a novel layered material, which has been most recently fabricated under high pressure (*Phys. Rev. Lett. 126(2021), 175501*). As a new class of 2D materials, in this work we conduct first-principles calculations to examine the stability and explore the electronic nature of MN$_4$ (M= Be, Mg, Ir, Rh, Ni, Cu, Au, Pd, Pt) monolayers. Acquired results confirm the dynamical and thermal stability of BeN$_4$, MgN$_4$, IrN$_4$, PtN$_4$ and RhN$_4$ monolayers. Interestingly, BeN$_4$ and MgN$_4$ monolayers are found to show anisotropic Dirac cones in their electronic structure. While PtN$_4$ monolayer is predicted to be a narrow band-gap semiconductor, IrN$_4$ and RhN$_4$ monolayers are found to be metallic systems. We also elaborately explore the effects of number of atomic layers on the electronic features of BeN$_4$ nanosheets, which reveal highly appealing physics. Our results highlight that BeN$_4$ nanosheet yield ultrahigh elastic modulus and mechanical strength, outperforming all other carbon-free 2D materials. Notably, RhN$_4$ nanosheet is predicted to yield high capacities of 562, 450 and 900 mAh/g, for Li, Na and Ca ions storages, respectively. This study provides a comprehensive understanding on the intrinsic properties of MN$_4$ nanosheets and highlight their outstanding physics.

Corresponding authors: *bohayra.mortazavi@gmail.com; **fshojaei@pgu.ac.ir
[#]These authors contributed equally

## 1. Introduction

Graphene [1–3], the most stable two-dimensional (2D) form of carbon atoms is proven to show exceptionally high thermal conductivity [4,5], carrier mobility and mechanical strength [6] and very appealing electronic and optical properties [7–10]. Carbon has two neighboring elements in the periodic table, boron and nitrogen. Borophene nanosheets, which are the 2D forms of boron atoms, have been recently fabricated by the epitaxy growth over silver substrate [11,12]. Borophene lattices likely to graphene do not show a band gap in their electronic structure and are metallic in their pristine form. Borophene nanosheets like their carbon counterpart are proven to show very high mechanical properties [13]. Since the most stable form of nitrogen atoms in ambient condition is a gas ($N_2$), unlike the boron and carbon, the 2D from of nitrogen atoms is extremely complex for fabrication. However, to date diverse 2D nanosheets consisting of nitrogen atoms have been experimentally synthesized. Particularly, numerous carbon-nitride nanosheets have been fabricated, such as: full-triazine $C_3N_3$ [14], holey $C_2N$ [15], graphene-like $C_3N$ [16], single-triazine $C_3N_4$ [17] and triazine $C_3N_5$ [18]. In contrast to the semimetallic nature of graphene, the aforementioned carbon-nitride nanosheets are semiconductors, which boost their application in nanoelectronics, optics and catalysis.

Despite of extensive and ongoing advances in the experimental realization of various 2D systems, nitrogen-rich lattices remain less explored. In a latest exciting advance by Bykov and coworkers [19], they devised a novel high-pressure approach and succeeded in the realization of $BeN_4$ beryllium polynitrides, a novel nitrogen-rich 2D structure. This latest accomplishment can be promising for the synthesis of a novel class of 2D materials with a chemical formula of $MN_4$, where M is a metal. Motivated by the experimental realization of $BeN_4$ nanosheet by Bykov *et al.* [19], in this study we examine the thermal and phononic stability, mechanical and electronic features of nitrogen-rich $MN_4$ (M= Be, Mg, Ir, Rh, Ni, Cu, Au, Pd, Pt) monolayer, using density functional theory (DFT) calculations. We particularly examine the application of these nanosheets as anodes for Li, Na or Ca-ion storage. Acquired results first-principles results provide a very comprehensive vision on the intrinsic properties of these novel 2D sheets and may serve as valuable guide for the future studies.

## 2. Computational methods

DFT calculations were performed with the generalized gradient approximation (GGA) and Perdew–Burke–Ernzerhof (PBE) [20], as implemented in *Vienna Ab-initio Simulation Package* [21,22]. Projector augmented wave method was used to treat the electron-ion interactions [23,24]. For the all considered structures we set a cutoff energy of 500 eV for the plane waves. For the geometry optimizations, atoms and lattices were relaxed according to the Hellman-Feynman forces using conjugate gradient algorithm until atomic forces drop to lower than 0.001 eV/Å [25]. The first Brillouin zone (BZ) was sampled with 12×12×1 Monkhorst-Pack [26] k-point grid. Mechanical properties are examined by conducting uniaxial tensile simulations over rectangular unitcells. Since PBE systematically underestimates the electronic band gaps, HSE06 hybrid functional [27] is employed to more precisely examine the electronic nature. DFT-D3 [28] van der Waals dispersion correction is considered for the modeling of multilayered structures and examination of adatoms adsorption. Moment tensor potentials (MTPs)[29] are trained to evaluate the phononic properties [30] using the MLIP package [31]. Ab-initio molecular dynamics (AIMD) simulations were conducted for 4×4×1 supercells with a time step of 1 ps [30]. The training sets for the development of MTPs are prepared by conducting two separate AIMD simulations at 50 K and from 100 to 700 K for 1000 and 1500 time steps, respectively. From every separate calculation, 500 trajectories were subsamples and used in the final training set. The phonon dispersions and group velocities are obtained for rectangular unitcell using the PHONOPY code [32] with fitted MTPs for interatomic force calculations [30] over 6×3×1 supercells. AIMD simulations are conducted at 500 K for 20000 time steps over supercells with 80 atoms to examine the thermal stability.

## 3. Results and discussions

In this section, we first introduce the crystal and structural features of MN$_4$ monolayers. Fig. 1a shows different views of BeN$_4$ monolayer's primitive unitcell and its corresponding in-plane lattice vectors ($\vec{a}$ and $\vec{b}$). At first glance, it can be seen that BeN$_4$ monolayer is one-atom thick and it is topographically flat. This material shows a rhombic primitive cell with lattice parameters of $|\vec{a}|$= 3.66 Å, $|\vec{b}|$= 4.27 Å and γ = 64.64° and P2/m (No. 10) space group. BeN$_4$ monolayer consists of armchair-shaped polymeric nitrogen chains (N$_\infty$) which are held together by Be atoms from their two sides, resulting in an anisotropic network exclusively made of pentagons (BeN$_4$) and hexagons (Be$_2$N$_6$). In obtained BeN$_4$, each N is triply coordinated (to two N atoms and one Be atom), while Be exhibits a slightly distorted square-planar coordination (to four pyridinic-like N atoms). In N$_\infty$, N-N bond length ($l_{N-N}$) is found to be about 1.34 Å which is an intermediate between length of single N-N (1.45 Å) and double N=N (1.21 Å) bonds. Our calculated lattice parameters and bond lengths are in excellent agreements with those in the original experimental work for BeN$_4$ [19]. From inorganic chemistry we know that a few transition metal ions also favor the coordination geometry of Be in BeN$_4$ (square planar coordination), namely, Ni, Cu, Pd, Rh, Ir, Pt, and Au atoms. Moreover, Mg may also sustain in this coordination environment. We constructed MN$_4$ (M = Mg, Ni, Cu, Pd, Rh, Ir, Pt, Cu, Au) monolayers by the substitution of Be with M atom and then carefully conducted the geometry optimization for the lattice parameters and ions positions. Among the considered lattices, AuN$_4$ distorted and ended up deforming to a different coordination environment, while the general feature of N$_\infty$ and square planar coordination was fully retained for other compositions. Spin-polarized calculations confirm that MN$_4$ (M = Mg, Ni, Cu, Pd, Rh, Ir, Pt, Cu) nanosheets possess non-magnetic ground states. The structural properties of these monolayers are summarized in Table 1. We performed Bader charge analysis for MN$_4$ monolayers to examine the nature of interaction between N$_\infty$ and M atoms. As summarized in Table 1, in all cases M atom is positively charged and the net transferred charge (metal-to- N$_\infty$) is almost equally distributed over N atoms. This leads to strong electrostatic interaction between the N$_\infty$ and M atoms. Interestingly, the amounts of transferred charge in BeN$_4$ and MgN$_4$ monolayers (1.67 and 1.62 $e$/cell) are by more than two times greater than those in transition metal based MN$_4$ monolayers. We also calculated the electron localization function (ELF) [33] to further analyze the nature of M-N$_\infty$ bonds in MN$_4$. The ELF results with isosurface value of 0.7 are depicted in Fig. 1(b). ELF takes a value between 0 to 1 at each point of the space. From Fig. 1(b), one can clearly see that electrons are strongly localized on N$_\infty$, in between each pair of neighboring N atoms (a $\sigma$ bond and a weak $\pi$ bond) and also on each N (lone pair electrons) pointing toward the M atoms. The small ELF values (< 0.01) in regions in between M and N atoms in transition metal-based MN$_4$ monolayers clearly indicates the absence of strong covalent interactions. However, in BeN$_4$ and MgN$_4$ monolayers, the electron could of anionic N$_\infty$ is strongly distorted and pulled toward the Be and Mg atoms, resulting in much higher covalent character of Be-N and Mg-N bonds. This may be attributed to the large positive Bader charge of Be and Mg atoms in BeN$_4$ and MgN$_4$ as well as their much smaller ionic radius compared to the other six transition metals.

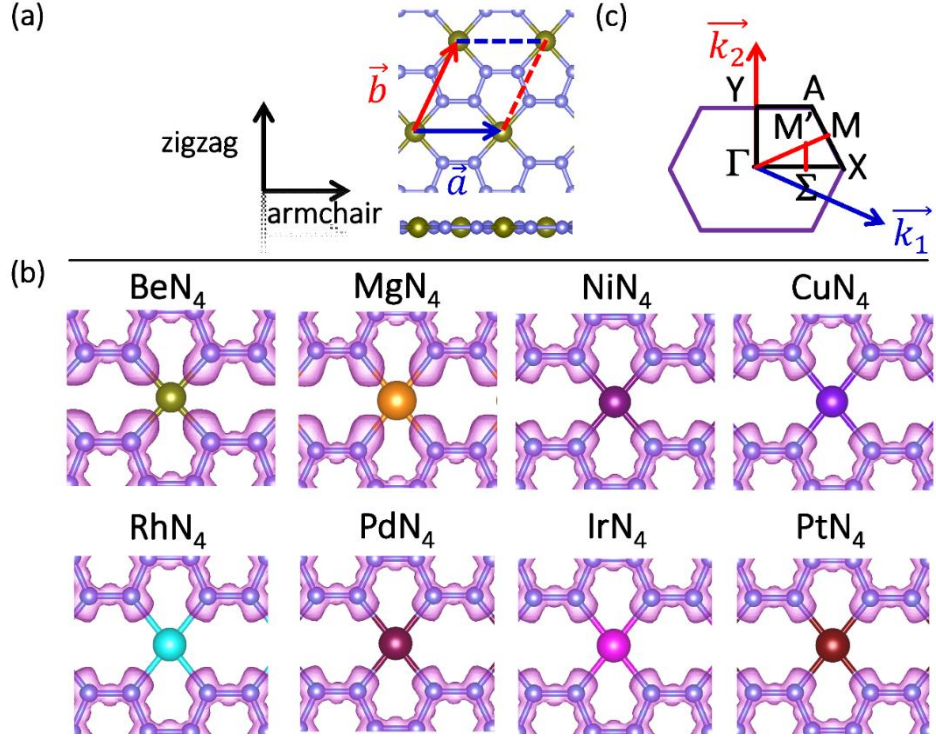

Fig. 1, (a) Top and side views of the crystal structure of BeN$_4$ monolayer. BeN$_4$ primitive cell and its in-plane lattice vectors are also shown. In figure, blue and green circles represent nitrogen (N) and beryllium (Be) atoms. (b) Isosurfaces of the electron localization function (ELF) calculated for MN$_4$ monolayers with M= Be, Mg, Ni, Cu, Rh, Pd, Ir, and Pt. For each one the isovalue is set to 0.7 (c) The corresponding first Brillouin zone (BZ), reciprocal lattice vectors, and high symmetry points along which the band structures are plotted.

Table 1, Computed structural, electronic, and dynamical properties of MN$_4$ nanosheets. All lengths are in Å and angles are in degree (°)

| System | ($|\vec{a}|, |\vec{b}|, \gamma$)[a] | $l_{N-N}$ | $l_{M-N}$ | q(M)[b] | Dynamical stability |
|---|---|---|---|---|---|
| BeN$_4$ | 3.66, 4.27, 64.64 | 1.343, 1.338 | 1.748 | 1.67 | √ |
| MgN$_4$ | 3.86, 4.88, 66.69 | 1.349, 1.355 | 2.057 | 1.62 | √ |
| NiN$_4$ | 3.72, 4.84, 65.49 | 1.332, 1.349 | 1.857 | 0.84 | × |
| CuN$_4$ | 3.81, 4.72, 66.75 | 1.322, 1.331 | 2.000 | 0.82 | × |
| RhN$_4$ | 3.77, 4.73, 66.52 | 1.336, 1.355 | 1.972 | 0.68 | √ |
| PdN$_4$ | 3.81, 4.76, 66.47 | 1.330, 1.349 | 2.002 | 0.72 | × |
| IrN$_4$ | 3.76, 4.74, 66.60 | 1.342, 1.366 | 1.964 | 0.78 | √ |
| PtN$_4$ | 3.79, 4.73, 66.39 | 1.337, 1.359 | 1.974 | 0.78 | √ |

[a] PBE optimized lattice parameters, lattice constants $|\vec{a}|$ and $|\vec{b}|$ and angle γ which is defined as the angle between lattice vectors $\vec{a}$ and $\vec{b}$. [b] Bader charge of M atom in MN$_4$ in unit of charge of an electron. [d] √ sign indicates that no band with imaginary frequency appear in the phonon band structure of MN$_4$ nanosheet, implying that the material is expected to be dynamically stable, while × sign indicates the presence of phonon modes with imaginary frequencies that can cause dynamical instability.

After describing the structural and bonding mechanism in MN$_4$ monolayers, we then examine dynamical and thermal stability by calculating the phonon dispersion relations. The acquired phonon dispersion relations for the considered monolayers are illustrated in Fig. 2. Like other 2D materials, all these monolayers show three acoustic modes starting from the Γ point, two with linear dispersions and the other one with quadratic relation. As it is conspicuous for the cases of MN$_4$ (N=Be, Mg, Ir, Rh and Pt) monolayers, the phonon dispersion relations are free of imaginary frequencies and thus confirming their desirable dynamical stability. Nonetheless, PdN$_4$, NiN$_4$ and CuN$_4$ monolayers show remarkable imaginary frequencies and are therefore dynamically unstable. For the dynamically stable lattices, we also assess the thermal stability by performing AIMD calculations at 1000 K for 20 ps. Results illustrated in Fig. S1 reveal that for MN$_4$ (N=Be, Mg, Ir, Rh and Pt) monolayers the total potential energies fluctuate around certain average values during the entire AIMD simulation, suggesting their outstanding thermal stability at the elevated temperature of 1000 K. Therefore, it can be concluded that among the considered lattices MN$_4$ (N=Be, Mg, Ir, Rh and Pt) nanosheets show desirable thermal and dynamical stability.

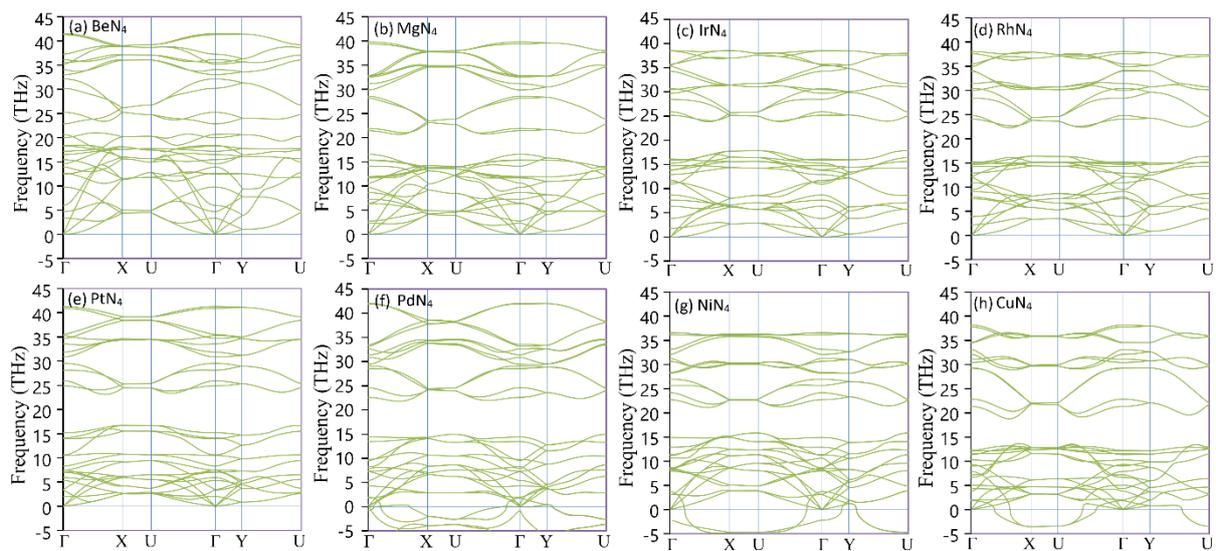

Fig. 2, Phonon dispersion relations of MN$_4$ monolayers acquired for rectangular unitcells.

We next elaborately examine the electronic properties of considered MN$_4$ monolayers. Worthy to note that it has been theoretically predicted in Ref. [19] that BeN$_4$ monolayer is semimetal with an anisotropic Dirac cone at the Fermi level. It is therefore highly interesting to investigate the evolution of Dirac cone in the electronic structure of MN$_4$ monolayers. For this goal we use PBE, mBJ,[34,35] and HSE06 [27] functionals to calculate the electronic band structures of all MN$_4$ monolayers along the high symmetry points in their Brillouin zone (BZ) (shown in Fig. 1c). We should note that mBJ functional is known to give band gaps with the same level of accuracy to HSE06 functional or GW methods, but with lower computational expenses [35]. Among $MN_4$ monolayers, $CuN_4$, $RhN_4$, and $IrN_4$ exhibit clear metallic character, expectable due to the odd number of electrons in their primitive cells. $NiN_4$ and $PdN_4$ monolayers that are earlier found to be dynamically unstable, also exhibit metallic behavior. However, our band structure calculations using the three functionals predict that like $BeN_4$, $MgN_4$ and $PtN_4$ monolayers are also semimetals with Dirac cone band dispersions (Fig. 3 and Fig. S2). In the following, we first investigate the electronic structures of these three monolayers in more detail and then briefly discuss the electronic properties of dynamically stable $RhN_4$ and $IrN_4$ monolayers. In Fig. 3, for each of $BeN_4$, $MgN_4$ and $PtN_4$ monolayers, we depicted PBE band structure, projected density of states (PDOS), and charge density distributions of Dirac bands at four selected *k*-points in the vicinity of Dirac point. From the illustrated results, it is apparent that $BeN_4$ is a semimetal with Dirac cone band crossing at an off-symmetry k-point ($\Sigma$) between $\Gamma$ and X points ($\Sigma$= 0.394($\kappa_1$/2) +0.197($\kappa_2$/2)). The general features of band structures obtained using mBJ, HSE06, and PBE functionals are similar with a minor difference that the position of Dirac point (DP) in mBJ band structure ($\Sigma$ = 0.425($\kappa_1$/2) + 0.212($\kappa_2$/2)) is slightly mismatched from that predicted by PBE and HSE06 methods (see Fig. 3 and Fig. S2). The off-symmetry position of DP in $BeN_4$ is explained to be attributed to the lower lattice symmetry of $BeN_4$ monolayer compared to graphene [19]. The Dirac cone band dispersion is known to lead to an ultrahigh carrier mobility in graphene [36,37]. To examine the carrier mobility around the DP, on can simply evaluate the Fermi velocities from the band structure data using $v_F = \frac{1}{\hbar}\frac{\partial E(k)}{\partial k}$, where $\frac{\partial E(k)}{\partial k}$ is the slope of Dirac cone bands and $\hbar$ is the reduced Planck's constant. We notice that in $BeN_4$, the slopes of Dirac cone bands are different along $\Gamma$-$\Sigma$ ($\kappa_x$) and $\Sigma$-M' ($\kappa_y$) directions, leading to anisotropic carrier velocities (see Fig. 3b). The Fermi velocities (Dirac band slopes) are found to be $6.04\times10^5$ m/s (+25 eV/Å) and $7.46\times10^5$ m/s (-30.85 eV/Å) along $\kappa_x$ direction and $2.92\times10^5$ m/s (±12.06 eV/Å) along $\kappa_y$ direction, in which the maximum value is by about a factor of 1.5 smaller than that of the graphene ($1.1\times10^6$ m/s) [37]. Our calculated Fermi velocities are in very good agreement with those reported in Ref. [19] ($8\times10^5$ m/s along $\kappa_x$ direction and $3.06\times10^5$ m/s along $\kappa_y$ direction). The nature of Dirac cone bands, however, has not been discussed in Ref. [19]. According to the PDOS and projected band structure, the bands in energy range of -2 to 3 eV are almost dominantly contributed by N($p_z$) states with a minor contribution of $p_z$ orbitals of Be atoms (find Fig. 3a). In addition, our charge density distribution analysis shows that near the Fermi level, the positive slope band is solely contributed by N atoms, representing a π(N-N) state, while the negative slope band which is almost equally contributed b N and Be atoms exhibits a π*(N-N) state hybridized with a π(Be-N) state (find Fig. 3c). These observations suggest that similar to graphene, delocalized π and π∗ states are mainly responsible for the emergence of Dirac cone in BeN$_4$ monolayer.

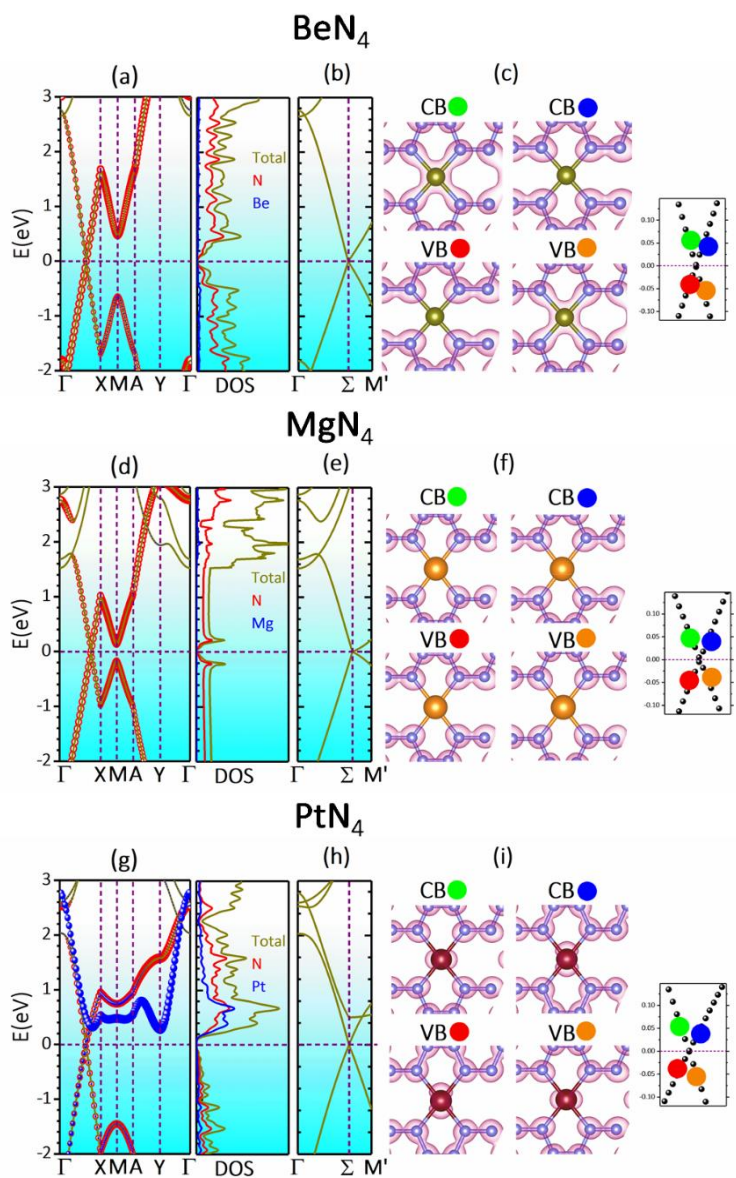

**Fig. 3**, PBE band structure and corresponding projected density of states (PDOS) as well as charge density distributions of Dirac bands at four selected *k*-points in the vicinity of Dirac point, calculated for BeN$_4$, MgN$_4$ and PtN$_4$ monolayers. In band structures red circles and blue filled dots indicate the contribution of N and metal atom to each band, respectively. Isosurface value of charge density distributions is set to 0.008 e/Å$^3$.

From the electronic band structures shown in Fig. 3, it is conspicuous that BeN$_4$ and MgN$_4$ monolayers share similarities in their electronic structures. This is an outstanding but also expected finding, since Be and Mg atoms belong to the same group and they thus show similar chemistry. Our careful investigation indicates that the aforementioned statements about the electronic properties of BeN$_4$ are also qualitatively consistent for the MgN$_4$ monolayer. However, in the following, we briefly highlight the two major differences that differentiate MgN$_4$ from BeN$_4$ monolayer. The first observation is that the anisotropic character of Dirac cone is more pronounced in MgN$_4$ monolayer (Fig. 3e). The estimated Fermi velocity (Dirac band slopes) of 6.70×10$^5$ m/s (-27.71 eV/Å) along the $\kappa_x$ direction is by more than 6 times larger than that along $\kappa_y$ direction (1.07 ×10$^5$ m/s (-4.42 eV/Å)), implying that Dirac fermions mostly transport long the armchair direction. This suggests that MgN$_4$ monolayer may be a promising candidate for applications in direction dependent quantum devices. The second major difference is that unlike the BeN$_4$, both Dirac cone bands in MgN$_4$ monolayer are exclusively made of N(p$_z$) states with no appreciable contribution from Mg states (find Fig. 3d and 3f). On the other hand, the electronic structure of PtN$_4$ monolayer is somehow different than MgN$_4$ and BeN$_4$ counterparts. As displayed in Fig. 3, PtN$_4$ is also a semimetal with Dirac cone band crossing in between $\Gamma$ and X points of its BZ ($\Sigma$ = 0.383($\kappa_1$/2) + 0.192($\kappa_2$/2)). The material also possesses an anisotropic Dirac cone, but its Fermi velocities (Dirac band slopes) of 5.25×10$^5$ m/s (+21.72 eV/Å) and 8.04×10$^5$ m/s (-33.27 eV/Å) along $\kappa_x$ direction and 5.43 ×10$^5$ m/s (±22.46 eV/Å) along $\kappa_y$ direction, are larger than the corresponding ones in MgN$_4$ monolayers. The charge density distributions of Dirac cone bands indicate that both positive and negative slope bands are derived from $\pi$ and $\pi*$ states of N$_\infty$ weakly hybridized with mostly d$_{yz}$ states of Pt atoms (find Fig. 3e). This further supports the idea that the $\pi$–conjugation system of N$_\infty$ play the key role in emerging Dirac cone in this class of materials. One may now raise this question that why do NiN$_4$ and PdN$_4$ monolayers with even number of electrons in their primitive cell exhibit metallic character? As shown in Fig. 4, the Dirac cones in band structure of NiN$_4$ and PdN$_4$ monolayers are by about 0.9 eV above the Fermi level and another band which is dominantly made of Pt(d$_{yz}$) states near the Fermi level, is partially occupied. This obviously indicates charge transfer form $\pi$ and $\pi*$ states of N$_\infty$ to the d$_{yz}$ orbitals of Ni and Pd atoms. Compared with that in PtN$_4$ monolayer, since d$_{yz}$ orbitals of Pt atoms has much higher energies than that of N and Pt, no effective charge transfer $\pi/\pi*$ to d$_{yz}$ charge transfer occurs in PtN$_4$ and the material stays a semimetal with a Dirac cone. Due to the fact that Pt is a rather heavy element and it also weakly contributes to the Dirac bands, the inclusion of spin-orbit coupling (SOC) should be taken into account for the examination of electronic response in PtN$_4$ monolayer. Interestingly, as shown in Fig. S3 we found that by incorporation of SOC effect in our PBE calculations, a marginal 0.06 eV band gap opens in the PtN$_4$ monolayer. As expected, the HSE06+SOC and mBJ+SOC methods predict slightly larger band gaps of 0.11 and 0.18 eV, respectively. This band gap is sufficiently large for practical applications in high-speed nanoelectronics. Fig. 4 also displays the band structure of RhN$_4$ and IrN$_4$, the two-of-five dynamically stable MN$_4$ monolayers. They both are metal and their band structures are almost similar with a Dirac cone at about 2 eV above the Fermi level. In both cases another Dirac cone exists at 0.7 eV below the Fermi level. However, since they are far from the Fermi level, they may not be practically useful as those with Dirac cone near the Fermi level.

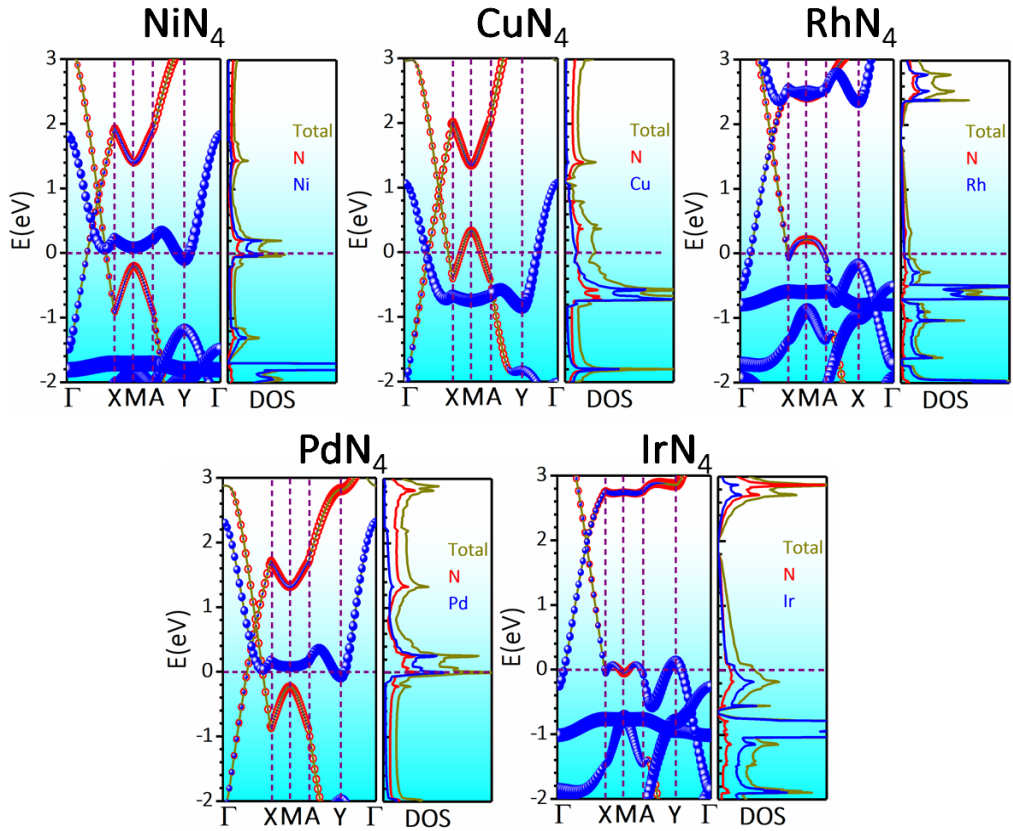

**Fig. 4**, PBE band structure and corresponding projected density of states (PDOS) calculated for NiN$_4$ and CuN$_4$, RhN$_4$ and PdN$_4$, IrN$_4$ monolayers. In band structures red circles and blue filled dots indicate the contribution of N and metal atom to each band, respectively.

Since BeN$_4$ is the most prominent member of this novel class of 2D materials and has been also experimentally fabricated in multilayer form, we now briefly examine the thickness dependent electronic properties of BeN$_4$ nanosheets. We used the experimentally resolved stacking order to construct bilayer to four-layer BeN$_4$ (2L- to 4L-BeN$_4$). This stacking order in 2L-BeN$_4$ can be simply generated by sliding one of layers of an AA-stacked configuration (by about 1.40 Å) along any of Be-N bonds (Find Fig. S4). After the energy minimization of multilayer BeN$_4$ systems, each two adjacent layers show the same stacking pattern as in 2L-BeN$_4$, which is in accordance with the bulk BeN$_4$ system. Fig 5, depicts the PBE band structures of 2L- to 4L-BeN$_4$ systems. For 2L-BeN$_4$, our careful investigation of states near the Fermi level indicates that due to the interlayer interactions Dirac cone states at Σ of the two monolayers hybridize resulting in two gapped Dirac cones (D$_1$ and D$_2$), one below and the other above the Fermi level. This band energy feature indicates p-type and n-type self-doping character of 2L-BeN$_4$. The existence of self-doping feature helps on achieving ultrahigh-speed holes in Dirac cone materials [38]. A similar character is also observed in 3L-BeN$_4$ and 4L-BeN$_4$. This makes multilayer BeN$_4$ systems also promising for high-speed nanoelectronics.

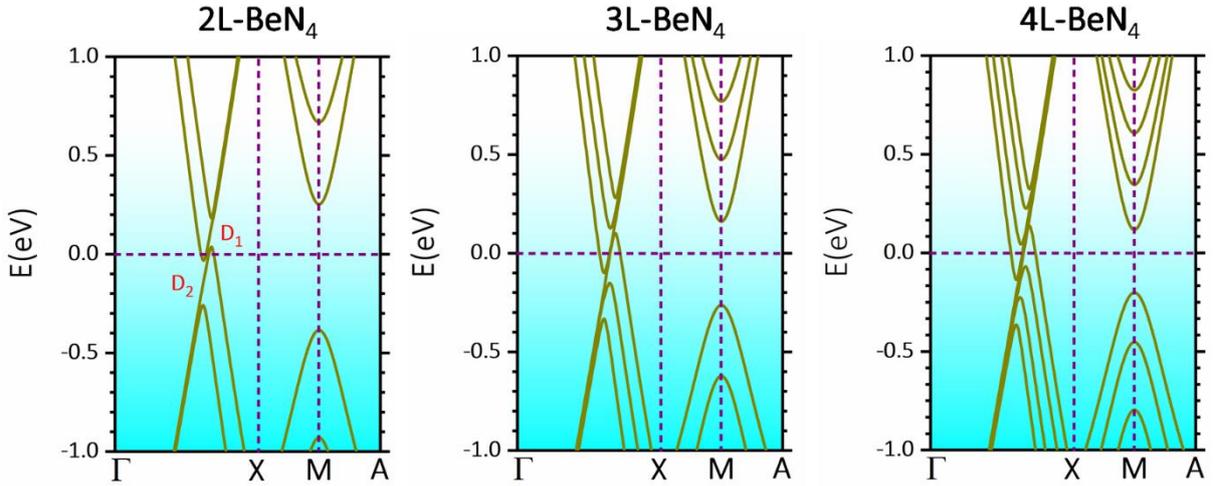

**Fig. 5**, PBE band structures of 2L- to 4L-BeN$_4$ nanosheets.

We now examine the feasibility of isolation of a single-layer BeN$_4$ from its multilayered or bulk counterparts by calculating the exfoliation process's cleavage energy. To evaluate this property, we consider a six-layered BeN$_4$ (6L-BeN$_4$) as illustrated in Fig. S4. After the complete geometry optimization of aforementioned structure, the last layer was gradually detached toward the vacuum with a faced step of 0.25 Å and at every step the changes in the energy with respect to the original configuration was recorded. The acquired results for the separation of a BeN$_4$ monolayer from the native single-layered structure is shown in Fig. 6. These results show that exfoliation energy increases sharply initially, but shortly after it reaches a plateau and after the distance of ~5 Å the van der Waals interactions with the detached monolayer vanishes convincingly. Our DFT-D3 [28] based simulations estimate a cleavage energies of 0.32 J/m$^2$, which is interestingly smaller than the experimentally measured cleavage energy of 0.37

J/m$^2$ for graphene [39]. This finding confirms the feasibility of the isolation of thermally and dynamically stable BeN$_4$ monolayer.

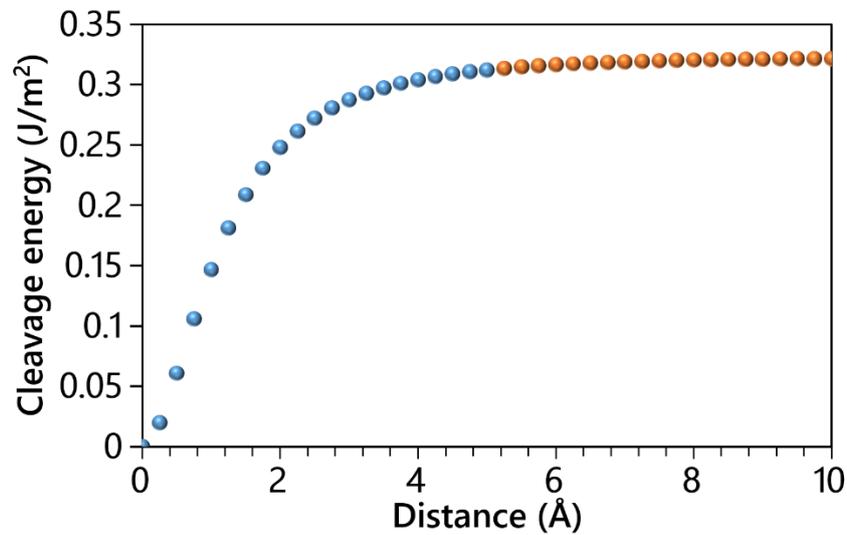

**Fig. 6**, Cleavage energy as a function of separation distance of BeN$_4$ monolayer.

We then investigate the mechanical properties of thermally and dynamically stable MN$_4$ (N=Be, Mg, Ir, Rh and Pt) monolayers by evaluating the uniaxial stress-strain relations. In these calculations the stresses along the two perpendicular directions of the loading are ought to stay negligible during various stages of the loading. Due to the contact with vacuum along the normal direction of the monolayers, the stress along this direction automatically reaches to a negligible value upon the geometry optimization. For the other planar direction, the periodic box size is altered to ensure the negligibility of stress. The mechanical responses are evaluated along the armchair and zigzag direction as distinguished in Fig. 1, in order to examine the anisotropy. The predicted uniaxial stress-strain relations of MN$_4$ (N=Be, Mg, Ir, Rh and Pt) monolayers are depicted in Fig. 7, in which we assumed a thickness of 3.06 Å [19]. Likely to the conventional materials, the uniaxial stress-strain curves show initial linear relations associated with the linear elasticity, followed by a nonlinear trend up to the maximum tensile strength point. The elastic modulus of BeN$_4$, MgN$_4$, IrN$_4$, RhN$_4$ and PtN$_4$ along the armchair(zigzag) directions are estimated to be 946(590), 582(353), 722(722), 711(615) and 687(652) GPa, respectively. Noticeably, the Poison's ratio of BeN$_4$ monolayer was predicted to be remarkably low and around 0.01. The maximum tensile strength of BeN$_4$, MgN$_4$, IrN$_4$, RhN$_4$ and PtN$_4$ along the armchair(zigzag) directions are found to be 100(43), 74(39), 76(71), 76(53) and 75(44) GPa, respectively. In Fig. 7a for the case of BeN$_4$ monolayer, we also plot the stress-train relation of graphene along the armchair direction, acquired using the same modeling procedure. Worthy to note the elastic modulus of graphene is found to be isotropic and around 1000 GPa, but the tensile strength along the armchair and zigzag are found to be 113 and 103 GPa, respectively. These results shows that the maximal elastic modulus and tensile strength of $BeN_4$ are remarkably close to those of the single-layer graphene. It can be concluded that $BeN_4$ record the highest elasticity and mechanical strength among the carbon-free 2D materials. In accordance with our earlier analysis on the electronic and bonding mechanism, $BeN_4$ and $MgN_4$ monolayers show very similar behavior, with highly anisotropic responses. Aforementioned nanosheets show almost twice higher tensile strengths along the armchair direction that the zigzag one. In contrast, $IrN_4$ monolayer exhibit convincingly isotropic mechanical response.

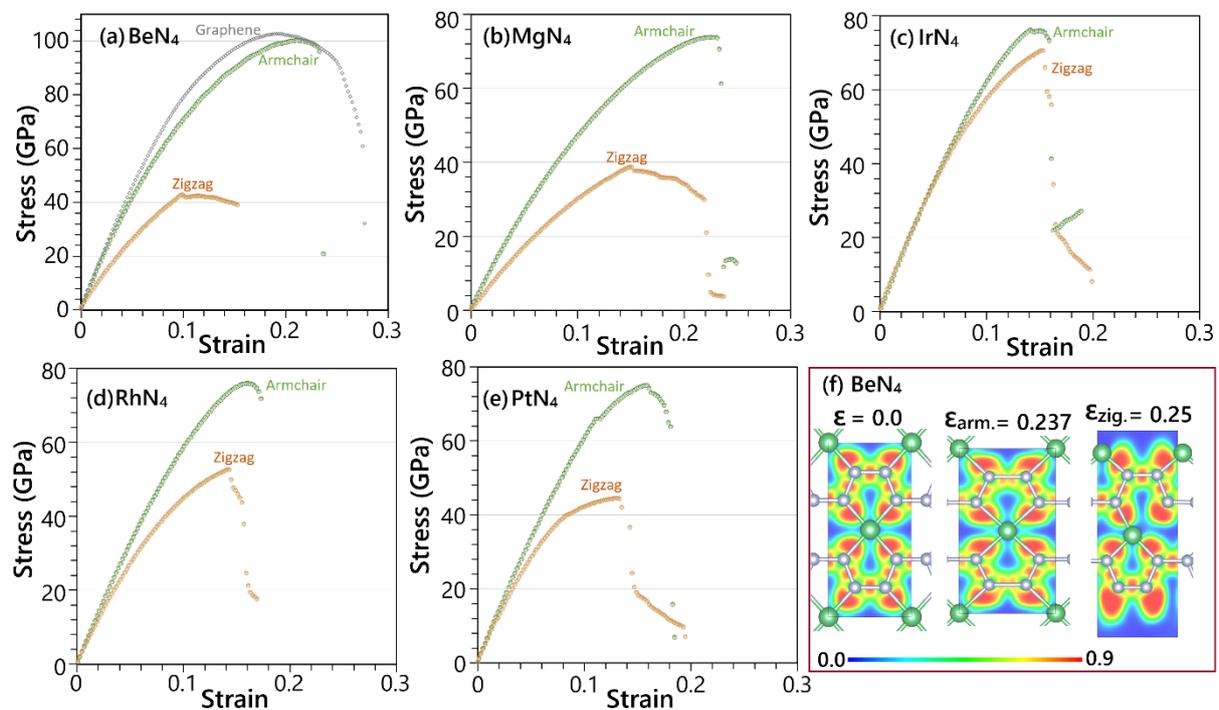

Fig. 7, Uniaxial stress-strain responses of (a) $BeN_4$, (b) $MgN_4$, (c) $IrN_4$, (d) $RhN_4$ and € $PtN_4$ along the armchair and zigzag directions. (f) Deformation process of $BeN_4$ monolayer at strain levels (ε) along the armchair (arm.) and zigzag (zig.) directions, in which contour illustrates the ELF.

Except for the case of $BeN_4$ monolayer, the tensile strength for all other considered nanosheets along the armchair direction are very close and around 75 GPa. To understand the anisotropic mechanical properties of considered nanomembranes, the mechanism of load transfer in should be taken into consideration. During the loading the bonds oriented along the loading directly involve in the stretching. As shown in Fig. 2f for the case of $BeN_4$ monolayer loaded along the armchair, at the strain levels close to the tensile strength, N-N bonds exactly oriented along the armchair direction show the maximum elongation. As it is apparent from the ELF contours, at high strain levels they start to split around the center of N-N bonds oriented along the loading, whereas for the rest of bonds in this systems they could keep their original pattern. The splitting of ELF contour is representative of the initiation of bond failure as the electron localization start to shift. This way, for the loading along the armchair direction the nature of N-N bonds in these systems play dominant behavior, which is not substantially affected by the type of metal atoms and thus lead to very close mechanical strengths and stretchability for very different metal atoms. In contrast for the loading along the zigzag direction, M-N bonds are oriented along the loading and directly get involved in the stretching and as shown for the case of BeN$_4$ monolayer in Fig. 2f, the fist rupture also occurs along these bonds. As discussed earlier, the nature of M-N bonds in these system can be remarkably affected by the change of metal atoms, which explains the remarkable variations for the mechanical properties along the zigzag direction for different compositions. The high elastic modulus of BeN$_4$ can be also indicative of a high lattice thermal conductivity in this novel 2D system. In Fig. 8 we therefore compare the phonon group velocities of BeN$_4$ with graphene. The maximum group velocity of acoustic modes in graphene and BeN$_4$ are predicted to be 21.9 and 19.9 km/s, respectively, which are very close. By comparing with MgN$_4$ and PtN$_4$, however a substantial decline in the phonon group velocities is observable. Lower symmetry and remarkable crossing of acoustic and optical bands in BeN$_4$ however suggest high phonon scattering rates in this system in comparison with graphene. Nonetheless, the predicted phonon group velocities suggest the high lattice thermal conductivity in BeN$_4$ nanosheets which can be an interesting topic for further studies. Presented results also reveal substantial suppression of phonon group velocities by increasing the weight of metal atoms in these systems.

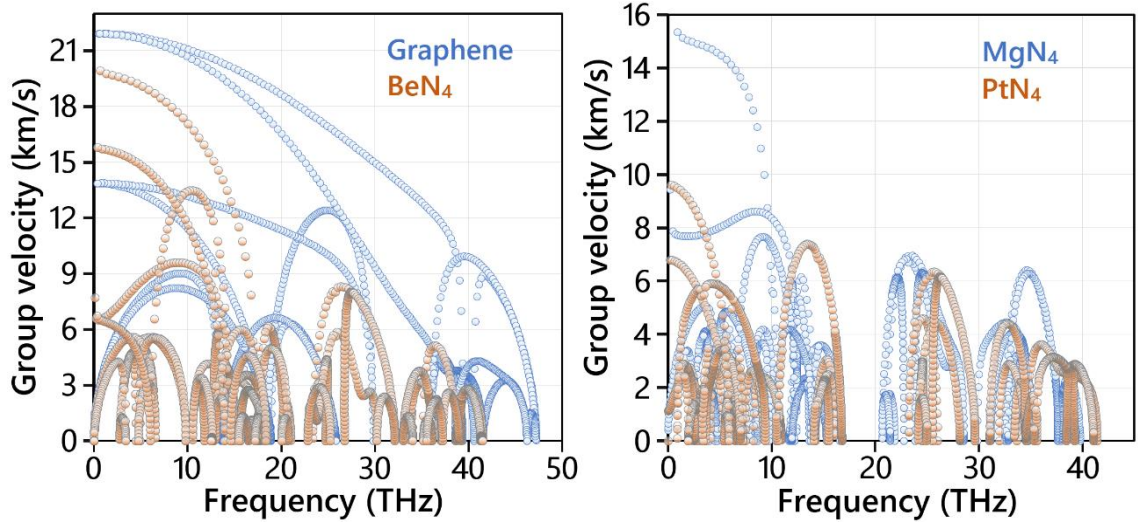

**Fig. 8**, Comparison of phonon group velocity of graphene, BeN$_4$, MgN$_4$, and PtN$_4$ monolayers calculated for primitive unitcells.

Light atomic weight and metallic electronic character of BeN$_4$, MgN$_4$ and RhN$_4$ can be promising for the application as anode materials in rechargeable metal ion-batteries. We therefore next examine the anodic properties of these nanosheets for the storage of Li, Na or Ca-ions. We first find the strongest binding site for different adatoms by comparing the adsorption energy. The adsorption energy, $E_{ad}$, is calculated via:

$$E_{ad} = E_{PA} - E_P - E_A \quad (1)$$

Here, $E_P$ is the total energy of MN$_4$ monolayer before the metal adatom adsorption, E$_{PA}$ is the total energy of the system after the adatom adsorption and E$_A$ is the per atom energy of the adatom in its most stable bulk structure [40]. In these simulations, for every adatom we considered five different original adsorption sites, as follows; over the metal atom, over the N atom, over the middle of N-N bonds or on the center of hexagonal or heptagonal rings. After the complete geometry optimization with conjugate gradient method, it was found that metal adatom either stabilize over the hexagonal or heptagonal rings. Results shown in Fig. 9 show the first two strongest binding sites of Li, Na or Ca adatoms over BeN$_4$, MgN$_4$ and RhN$_4$ monolayers and their corresponding adsorption energies. Bader charge analysis results reveal that for all monolayers upon the adsorption Li and Na adatoms they become fully ionized, whereas for the case of Ca atoms only one electron was transferred to the substrate. Results shown in Fig. 9 reveal positive adsorption energies for Li, Na or Ca adatoms over BeN$_4$ and MgN$_4$ nanosheets, which reveal the adatoms would prefer to form metal clusters rather than the adsorption over the surface. It is thus clear that BeN$_4$ and MgN$_4$ nanosheets are not desirable for the application as anodes in rechargeable metal ion batteries. As discussed earlier for the results shown in Fig. 3, in BeN$_4$ and MgN$_4$ monolayers the electronic states near the Fermi energy are mostly due to the N atoms rather than the metal atom. In contrast for the case of RhN$_4$ nanosheet, Pt atoms show remarkable contribution to the states around the Fermi level, which consequently enhance the binding with metal adatoms. As a result, Li, Na or Ca adatoms show negative adsorption energies over RhN$_4$ monolayer and confirm they prefer to adsorb over the surface rather than forming metallic clusters. Similar observation should be also consistent for the cases of IrN$_4$ and PtN$_4$ monolayers, however their higher atomic weights suppress the storage capacity.

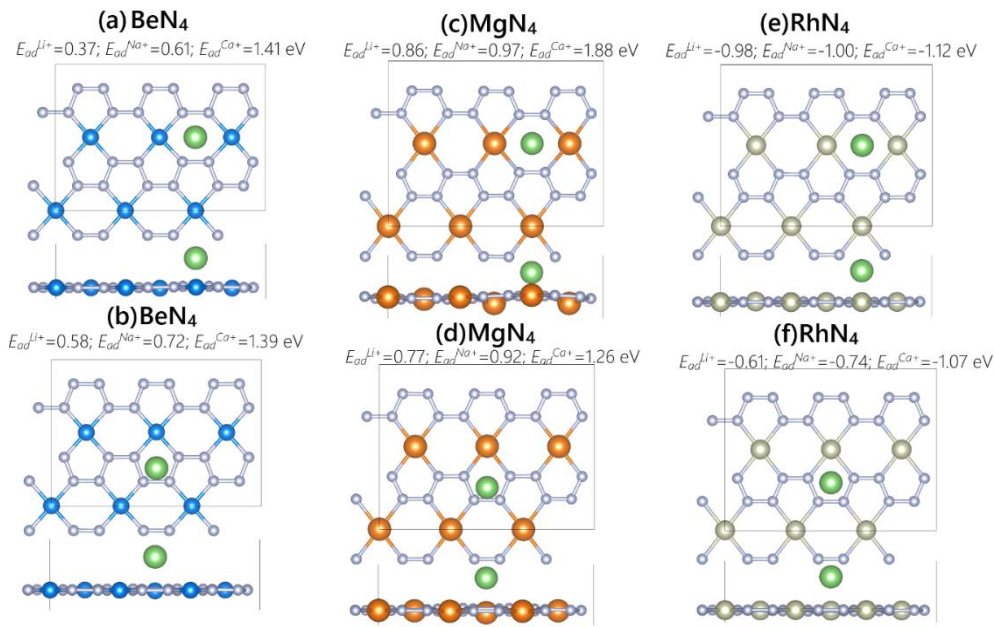

**Fig. 9**, Top and side views for the first two strongest binding sites of Li adatoms adsorption over (a, b) BeN$_4$, (c, d) MgN$_4$ and (e, f) RhN$_4$ monolayers. Corresponding adsorption energies ($E_{ad}$) for every adsorption site is mentioned for Li, Na and Ca ions.

At this point we increase the content of Li, Na and Ca adatoms on the both sides of RhN$_4$ monolayer. To this goal, first all the strongest adsorption sites are filled and next the second binding sites are filled. By increasing the adatoms coverage, we evaluate the average adsorption energy ($E_{av-ad}$) profile by:

$$E_{av-ad} = \frac{(E_{PA} - n \times E_A - E_P)}{n} \qquad (2)$$

Here, $E_{PA}$ is the total energy of RhN$_4$ monolayer with "$n$" adsorbed metal atoms. The average adsorption energy profiles for the Li, Na and Ca adatoms over RhN$_4$ monolayer are shown in Fig. 10. As it is clear, for the range of studied contents, the average adsorption energies stay negative, confirming that metal adatoms prefer to adsorb on the surface. For the cases of Na and Ca atoms we found that due to their larger ionic sizes they prefer to form the second layer on each side of RhN$_4$ after a certain limit, whereas more Li atoms can be adsorbed. The storage capacity is calculated by finding the maximum content of adatoms that can adsorb in the single-layer form on each side of the RhN$_4$ monolayer (as shown in Fig. 10). According to this concept, the storage capacity of RhN$_4$ nanosheet for Li, Na and Ca ions is predicted to be 562, 450 and 900 mAh/g, respectively. Predicted storage capacities are considerably higher than that of the commercial graphite for Li-ions, 372 mAh/g [40]. As expected, our results for the electronic density of states reveal that upon the adsorption of metal ions over the RhN$_4$ monolayer the metallic electronic character was fully preserved. Metallic nature in these systems can be representative of good electronic conductivity, which is highly desirable for the application in rechargeable metal ion batteries.

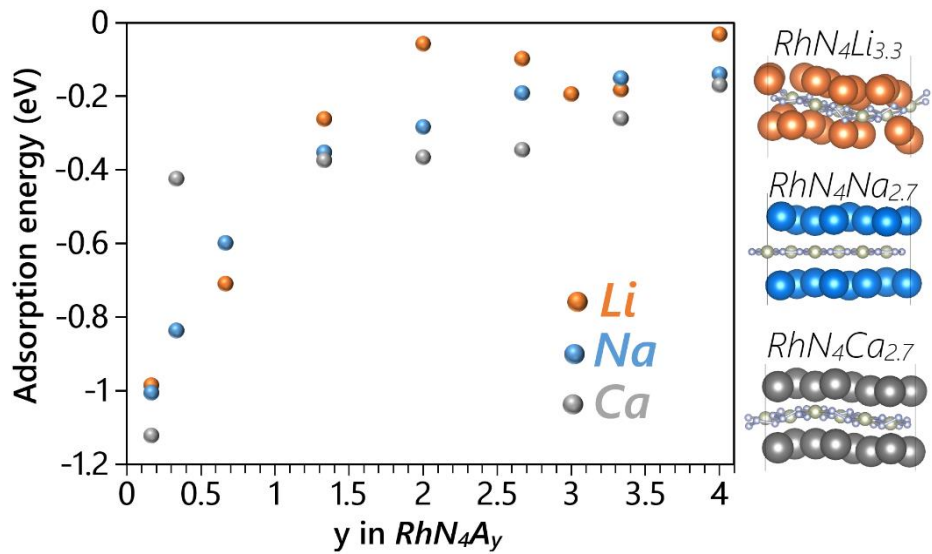

**Fig. 10,** Average absorption energy as a function of Li, Na, Ca ions content over RhN$_4$ monolayer. RhN$_4$ monolayer with maximum content of adatoms in the single-layer form on each side are shown in the right side.

4. Concluding remarks

Motivated by the latest experimental accomplishment in the synthesis of BeN$_4$ nanosheets under high pressure, herein first-principles calculations are conducted to examine the stability and intrinsic properties of MN$_4$ (M= Be, Mg, Ir, Rh, Ni, Cu, Au, Pd, Pt) monolayers. First-principles results confirm dynamical and thermal stability of BeN$_4$, MgN$_4$, IrN$_4$, PtN$_4$ and RhN$_4$ monolayers. Notably, BeN$_4$ and MgN$_4$ monolayers are found to show Dirac cones with anisotropic Fermi velocities. PtN$_4$ monolayer is however confirmed to show a band gap of 0.18 eV on the basis of mBJ+SOC method. IrN$_4$ and RhN$_4$ monolayers are predicted to exhibit metallic character in their pristine form. The effects of number of atomic layers on the electronic characteristics of BeN$_4$ nanosheets are also examined. We observed p-type and n-type self-doping character of multilayer BeN$_4$, which can be promising for high-speed nanoelectronics. Our results highlight that BeN$_4$ nanosheet yield ultrahigh elastic modulus and mechanical strength, outperforming all other carbon-free 2D materials. MgN$_4$, IrN$_4$, PtN$_4$ and RhN$_4$ nanosheets are found to exhibit outstandingly high maximal tensile strengths around 75 GPa. Moreover, BeN$_4$, MgN$_4$, PtN$_4$ and RhN$_4$ monolayers are found to show highly anisotropic mechanical response, with distinctly higher elastic modulus and tensile strength along the armchair that the zigzag direction. The cleavage energy of BeN$_4$ monolayer is predicted to be 0.32 J/m$^2$, which is smaller than that of graphene and confirms the feasibility of BeN$_4$ monolayer isolation form its multilayer counterparts. BeN$_4$ nanosheets is also predicted to show high phonon group velocities, close to those of graphene, which can be an indication of high lattice thermal conductivity. Substantial suppression of phonon group velocities is observed by increasing the wright of metal atoms in MN$_4$ monolayers. While BeN$_4$ and MgN$_4$ nanosheets are examined not to be suitable for application as anode materials in metal-ion batteries, RhN$_4$ lattice is predicted to yield remarkably high storage capacities of 562, 450 and 900 mAh/g, for Li, Na and Ca ions, respectively. Worthy to mention that the metal atom in MN$_4$ monolayers can also take other elements, such as Ag, Co, Fe and Co, which should be explored in the future studies. Our extensive first-principles results highlight the outstanding physics of MN$_4$ nanosheets and suggest them as promising candidates for nanoelectronics, structural components and energy storage/conversion systems.

## Acknowledgment

B.M. and X.Z. appreciate the funding by the Deutsche Forschungsgemeinschaft (DFG, German Research Foundation) under Germany's Excellence Strategy within the Cluster of Excellence PhoenixD (EXC 2122, Project ID 390833453). F.S. thanks the Persian Gulf University Research Council, Iran for support of this study. B. M is greatly thankful to the VEGAS cluster at Bauhaus University of Weimar for providing the computational resources.## Acknowledgment

B.M. and X.Z. appreciate the funding by the Deutsche Forschungsgemeinschaft (DFG, German Research Foundation) under Germany's Excellence Strategy within the Cluster of Excellence PhoenixD (EXC 2122, Project ID 390833453). F.S. thanks the Persian Gulf University Research Council, Iran for support of this study. B. M is greatly thankful to the VEGAS cluster at Bauhaus University of Weimar for providing the computational resources.

## Appendix A. Supplementary data

The following are the supplementary data to this article:

# Supplementary information

# Ultrahigh stiffness and anisotropic Dirac cones in BeN$_4$ and MgN$_4$ monolayers: A first-principles study

Bohayra Mortazavi [a,*], Fazel Shojaei[b], Xiaoying Zhuang[a,c][a]Chair of Computational Science and Simulation Technology, Institute of Photonics, Department of Mathematics and Physics, Leibniz Universität Hannover, Appelstraße 11,30167 Hannover, Germany.[b]Department of Chemistry, Faculty of Nano and Bioscience and Technology, Persian Gulf University, Bushehr 75169, Iran.[c]College of Civil Engineering, Department of Geotechnical Engineering, Tongji University, 1239 Siping Road Shanghai, China.
Bohayra Mortazavi [a,*], Fazel Shojaei[b], Xiaoying Zhuang[a,c]
[a]Chair of Computational Science and Simulation Technology, Institute of Photonics, Department of Mathematics and Physics, Leibniz Universität Hannover, Appelstraße 11,30167 Hannover, Germany.
[b]Department of Chemistry, Faculty of Nano and Bioscience and Technology, Persian Gulf University, Bushehr 75169, Iran.
[c]College of Civil Engineering, Department of Geotechnical Engineering, Tongji University, 1239 Siping Road Shanghai, China.

## 1. Atomic lattices in VASP POSCAR format.

**BeN4**
```
   1.00000000000000
     3.6599429756439656    0.1369011627511460    0.0000000000000000
     1.6846574069694418    3.9279213095315524    0.0000000000000000
     0.0000000000000000    0.0000000000000000   20.0000000000000000
   N    Be
     4     1
Direct
  0.1512106903615432  0.3405305751020802  0.5000000000000000
  0.5179513333843172  0.3405724162839941  0.5000000000000000
  0.4899664570818420  0.6632643226725143  0.5000000000000000
  0.8567060551059755  0.6633054228612695  0.5000000000000000
  0.0039581567167701  0.0019187456560132  0.5000000000000000
```

**MgN4**
```
   1.00000000000000
     3.8599048317633984    0.0788309043067702    0.0000000000000000
     1.8392663497583976    4.5195578927074189    0.0000000000000000
     0.0000000000000000    0.0000000000000000   20.0000000000000000
   N    Mg
     4     1
Direct
  0.1464347689995541  0.3658160416343179  0.5000000000000000
  0.4976809168614001  0.3656772824850069  0.5000000000000000
  0.5102382565009604  0.6381591097107515  0.5000000000000000
  0.8614804320224900  0.6380204428296423  0.5000000000000000
  0.0039583182660508  0.0019186059161527  0.5000000000000000
```

**CuN4**
   1.00000000000000
     3.8096867522426789    0.1253913370108510    0.0000000000000000
     1.7192661133312872    4.3953629556167977    0.0000000000000000
     0.0000000000000000    0.0000000000000000   20.0000000000000000
   N    Cu
     4     1
Direct
  0.1483781199294976  0.3637062238416409  0.5000000000000000
  0.4951938617886935  0.3638612685697668  0.5000000000000000
  0.5127234822762848  0.6399746869177504  0.5000000000000000
  0.8595389178905667  0.6401306858571995  0.5000000000000000
  0.0039583107653982  0.0019186173895065  0.5000000000000000

**IrN4**
   1.00000000000000
     3.7598997674777945    0.1481743290745947    0.0000000000000000
     1.7086121661174742    4.4179071427816643    0.0000000000000000
     0.0000000000000000    0.0000000000000000   20.0000000000000000
   N    Ir
     4     1
Direct
  0.1477719757375260  0.3576440882122667  0.5000000000000000
  0.5044677884168373  0.3576140523036077  0.5000000000000000
  0.5034495399500848  0.6462225193876874  0.5000000000000000
  0.8601451738978279  0.6461926323967280  0.5000000000000000
  0.0039582146481720  0.0019181902755818  0.5000000000000000

**NiN4**
   1.00000000000000
     3.7250489174759904    0.1416759369362366    0.0000000000000000
     1.7040469762438153    4.1477809703920245    0.0000000000000000
     0.0000000000000000    0.0000000000000000   20.0000000000000000
   N    Ni
     4     1
Direct
  0.1512196414400165  0.3500416737008624  0.5000130276234900
  0.5084919930324052  0.3499210215261903  0.4999984856605175
  0.4994258208952354  0.6539158195087131  0.5000015143394824
  0.8566971879517510  0.6537941789840173  0.4999869723765101
  0.0039580493310396  0.0019187888560879  0.5000000000000000

**PdN4**
```
   1.00000000000000
     3.8098797731380469    0.1467297609498282    0.0000000000000000
     1.7298437958702404    4.4304992379924499    0.0000000000000000
     0.0000000000000000    0.0000000000000000   20.0000000000000000
   N    Pd
   4    1
Direct
  0.1492187933277336   0.3620589880033904   0.5000000000000000
  0.4981960634878249   0.3620171685800955   0.5000000000000000
  0.5097210154979805   0.6418193738628659   0.5000000000000000
  0.8586983360618561   0.6417775093896125   0.5000000000000000
  0.0039584842750529   0.0019184427398926   0.5000000000000000
```

**PtN4**
```
   1.00000000000000
     3.7879132809859328    0.1480106721709665    0.0000000000000000
     1.7218437789351744    4.4005086942962981    0.0000000000000000
     0.0000000000000000    0.0000000000000000   20.0000000000000000
   N    Pt
   4    1
Direct
  0.1491804115998188   0.3588294377992172   0.5000000000000000
  0.5017600214191974   0.3588693529333714   0.5000000000000000
  0.5061573069477247   0.6449672187579236   0.5000000000000000
  0.8587367380355351   0.6450072828097771   0.5000000000000000
  0.0039582146481720   0.0019181902755818   0.5000000000000000
```

**RhN4**
```
   1.00000000000000
     3.7670809974945363    0.1481306396407278    0.0000000000000000
     1.7127896998326260    4.4110189829679598    0.0000000000000000
     0.0000000000000000    0.0000000000000000   20.0000000000000000
   N    Rh
   4    1
Direct
  0.1481511207031231   0.3592731268105692   0.5000000000000000
  0.5024206382633261   0.3592823158589472   0.5000000000000000
  0.5054964407224793   0.6445542265840145   0.5000000000000000
  0.8597660086864665   0.6445633705824336   0.5000000000000000
  0.0039584842750529   0.0019184427398926   0.5000000000000000
```

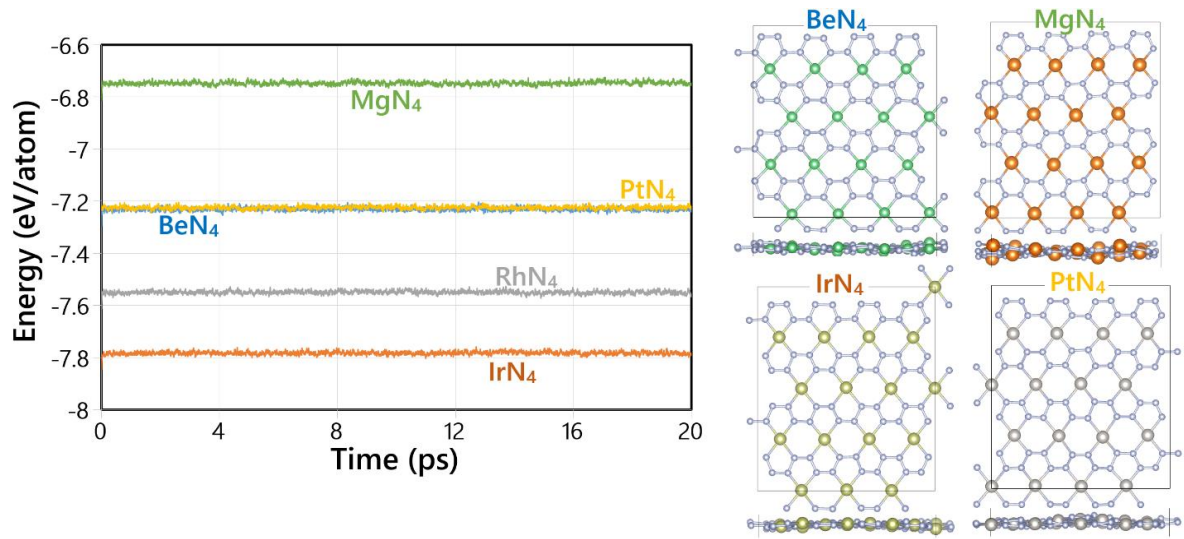

**Fig. S1**, (left) Fluctuation of potential energy of MN$_4$ (N=Be, Mg, Ir, Rh and Pt) monolayers and (right) top view of rectangular supercell after 20 ps AIMD simulations at 1000 K.

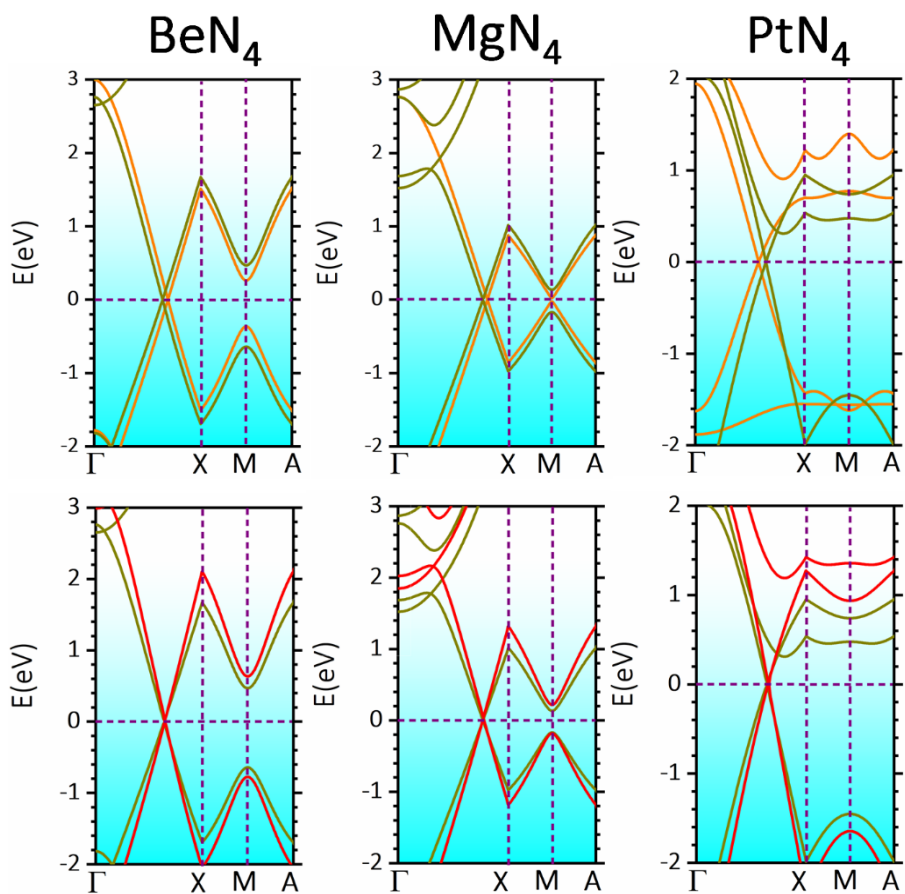

**Fig. S2**, Electronic band structures of BeN$_4$, MgN$_4$ and PtN$_4$ monolayers, calculated using mBJ (orange line) and HSE06 (red line) functional. For comparison, PBE band structure (green line) is also shown for each lattice.

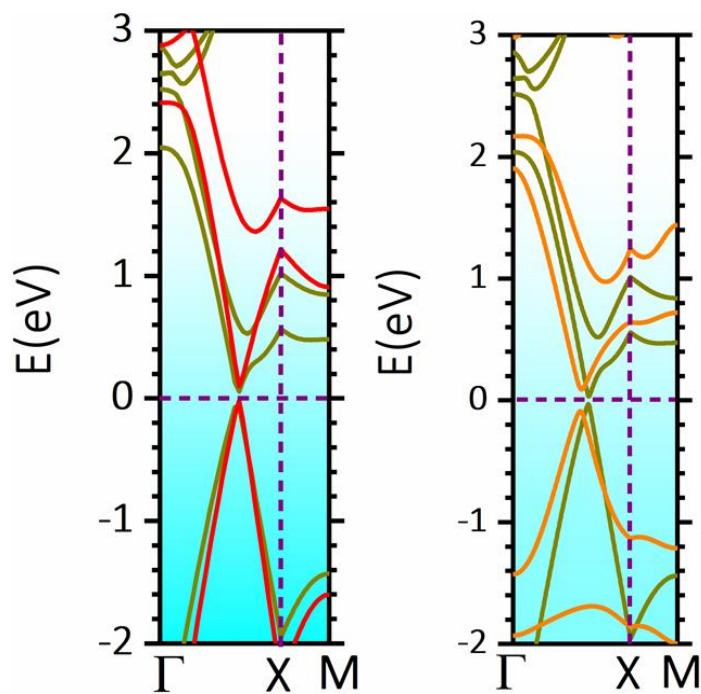

Fig. S3, Electronic Band structure of PtN$_4$ calculated using PBE+SOC (green line), HSE06+SOC (red line), and mBJ+SOC (red line) methods.

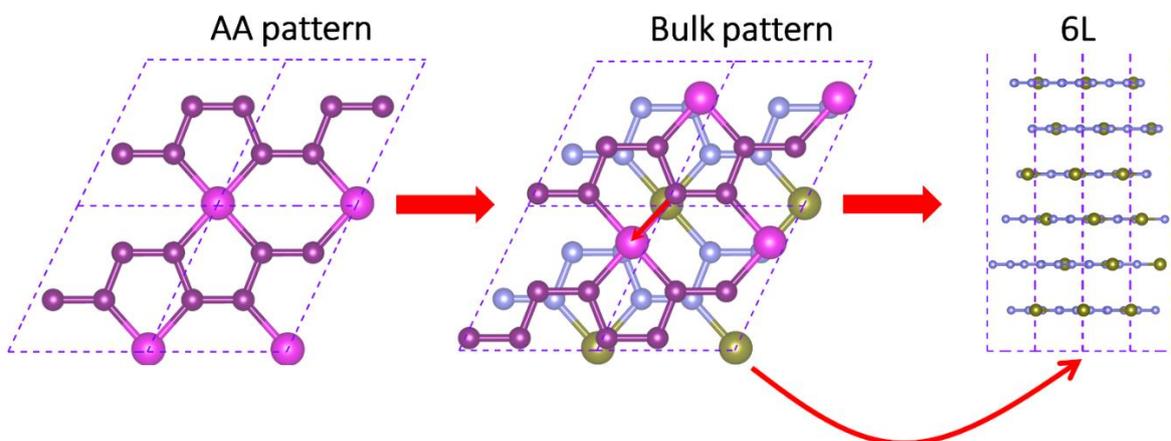

Fig. S4, Top view of crystal structure of 2L-BeN$_4$ in two stacking patterns of AA and BeN4 experimental (Bulk) pattern. The bulk stacking pattern in 2L-BeN$_4$ can be simply generated by sliding one of layers of an AA-stacked configuration (by about 1.40 Å) along any of Be-N bonds (shown by a thin red arrow). The side view of 6L-BeN$_4$ is also shown. In any multilayer BeN$_4$ system considered in this work, each two adjacent layers have the same stacking pattern as in 2L-BeN$_4$. In figure, blue and purple circles represent N atoms, while green and pink circles represent Be atoms.